# Efficient Beam Converter for the Generation of Femtosecond Vortices

Vladlen G. Shvedov[1,2],* Cyril Hnatovsky[1], Wieslaw Krolikowski[1], Andrei V. Rode[1]

[1]*Laser Physics Centre and* [2]*Nonlinear Physics Centre,*

*Research School of Physics and Engineering, The Australian National University,*

*Canberra ACT 0200, Australia*

*Corresponding author: vgs124@physics.anu.edu.au

**Abstract**: We describe an optical beam converter for an efficient transformation of Gaussian femtosecond laser beams to single- or double-charge vortex beams showing no spatial or topological charge dispersion. The device achieves a conversion efficiency of 75% for single- and 50% for double-charge vortex beams and can operate with high energy broad bandwidth pulses. We also show that the topological charge of a femtosecond vortex beam can be determined by analyzing its intensity distribution in the focal area of a cylindrical lens.

Over the last two decades femtosecond laser pulses have been extensively used both for precision materials processing and in fundamental studies of light-matter interactions. High intensity ultrashort optical vortex pulses provide an opportunity to investigate the effects of the angular momentum on atomic or molecular systems and transient non-equilibrium states of matter [1, 2]. Current methods of generating femtosecond optical vortices with spiral phase plates and holograms are inherently chromatic and therefore require the introduction of correcting elements in the attempt to compensate topological charge dispersion caused by the broad spectral bandwidth associated with ultrashort light pulses [3, 4].

We have recently demonstrated the formation of white light cw vortices using polarization singularities in birefringent crystals [5]. In this Letter we extend this idea to femtosecond pulses. In particular, we propose a compact polarization-singularity beam converter for the generation of single- and double-charge femtosecond laser vortex beams. The converter is: i) insensitive to the spectral width of the ultra-short pulses as, by definition, it has zero spatial and topological-charge dispersion; ii) yields high conversion efficiencies of 75% for single- and 50% for double-charge vortex beams; iii) is suitable for the operation with high energy pulses as it is nominally free of any light absorbing elements, operates only with divergent laser beams and uses standard high optical damage threshold optics (i.e., ~100 GW/cm$^2$); and iv) can be adapted with minimum adjustment for a wide range of wavelength and beam power conditions. It should be mentioned that a similar idea of using polarization singularities to generate polychromatic vortices has been recently implemented with an axially symmetric polarizer [6]. However, in that case only double-charge vortex pulses can be generated with a much lower conversion efficiency of 25%. Moreover, because of energy dissipation inside the axial polarizer (viz., 50% of the incident circularly polarized beam) the scheme cannot be used in a high pulse energy regime.

To synthesize femtosecond vortex beams we exploit polarization singularities which can be created when a beam of light propagates through an anisotropic medium [7, 8, 9, 10]. Polarization singularities can be aligned along the same or very close directions within a broad spectral range of the incident light field [11, 12]. In particular, it was shown in [8,9] that focusing a beam into a uniaxial crystal generates an optical vortex with the polarization singularity propagating along the optical axis of the crystal. A polarization singularity can be converted into a phase singularity by a polarization filter [9]. As a result, a point of zero intensity is maintained along the whole singularity line [10].

When a circularly polarized Gaussian beam enters a uniaxial crystal along its optical axis a superposition of two vector states of the field is generated after the crystal: i) a non-vortex state of the same handedness and ii) a double-charge optical vortex of the opposite handedness (Fig.1) [10]:

$$\mathbf{E}_\perp^\pm = \mathbf{c}^\pm (G_o + G_e) - \mathbf{c}^\mp \frac{1}{v^2}\left[\left(uv + w_0^2 \xi_o\right)G_o - \left(uv + w_0^2 \xi_e\right)G_e\right] \qquad (1)$$

where $\mathbf{E}_\perp$ denotes solutions of the paraxial wave equation in an anisotropic medium $(\nabla_\perp^2 + 2ikn_o \partial_z)\mathbf{E}_\perp = -\gamma \nabla_\perp (\nabla_\perp \mathbf{E}_\perp)$ for the transverse components of the electric field. In the above expressions: $\nabla_\perp \equiv \mathbf{e}_x \partial_x + \mathbf{e}_y \partial_y$; $\gamma = (n_o^2 - n_e^2)/n_e^2$, where $n_o$ and $n_e$ is the ordinary and extraordinary refractive index;

$G_{o;e} = \frac{E}{\xi_{o;e}} \exp\left(-\frac{uv}{w_0^2 \xi_{o;e}}\right)$; $\xi_o = 1 + 2i\frac{z}{k n_o w_0^2}$; $\xi_e = 1 + 2i\frac{z \eta}{k n_e w_0^2}$; $(u,v) = x \pm iy$ are independent spatial variables, $w_0$ is the beam waist, and $\mathbf{c}^\pm = (\mathbf{e}_x \pm i\mathbf{e}_y)/\sqrt{2}$ is the unit wave polarization vector. Here we note that the conversion of the incident wave into a vortex along the optical axis of a uniaxial crystal is wavelength independent and can therefore be achieved for any harmonic of the incident light within the transparency window of the crystal. This important property dramatically facilitates the generation of axially symmetric polychromatic femtosecond beams with zero intensity at their axes.

One can see from Eq.(1) that the power conversion efficiency of the incident beam into an optical vortex depends on both, the crystal (viz., optical thickness and birefringence) and the laser beam parameters: $P^\mp(z) = \left(1 - (1 + z^2/L^2)^{-1}\right)P^\pm(0)/2$, where $P^\mp(z)$ is the power of an optical vortex after travelling a distance $z$ along the axis of the crystal, $P^\pm(0)$ is the power of the incident circularly polarized Gaussian beam and $L = kw_0^2 n_e^2/(n_o^2 - n_e^2)$. The superscripts ($\pm$) denote the handedness of the circular polarization in the incident and the transformed beams. It also follows from the above expressions that the beam-vortex conversion efficiency increases with the crystal length and the divergence of the beam. The maximum achievable ratio of the optical powers in the double-charge vortex beam and the non-vortex beam is therefore 1:1 or 50% of the total power incident onto the polarisation beam splitter (PBS) shown in Fig.1.

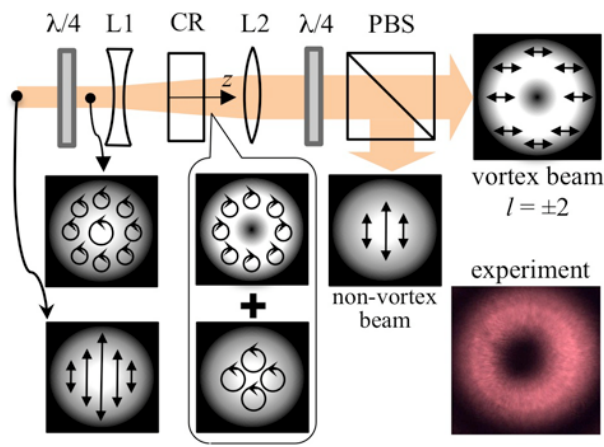

**Fig. 1.** Setup for the generation of double-charge femtosecond vortex pulses. $\lambda/4$ – achromatic quarter-wave plate, L1 – negative lens, CR – uniaxial crystal, L2 – positive lens, PBS – polarization beamsplitter. Polarization states after each optical element are indicated by arrows. Bottom right: CCD image (4.2 x 4.2 mm$^2$) of a double-charge vortex recorded ~5 meters after the converter.

In the previous studies [9, 11, 122] the generation of polychromatic (white) light vortices was achieved by focusing the incident beam inside the crystal or close to its surface. This approach is well-adapted for the creation of low intensity vortices, but is unsuitable for high energy femtosecond laser pulses due to possible super-continuum generation or optical breakdown inside the crystal. To avoid this problem, the proposed design operates only with divergent laser beams, which still ensures efficient light conversion (Fig. 1). Here, the circularly polarized femtosecond beam is defocused with a negative lens L1 and after propagating along the optical axis of a 10 mm-long $Ca_2CO_3$ crystal (in fact, any *c*-cut uniaxial crystal is suitable for this purpose) is collimated with a positive lens L2. The second quarter-wave plate together with a polarization beamsplitter PBS separate the emerging double-charge vortex from a non-vortex beam. In such a setup, the laser power is not dissipated inside optical elements of the system, which is very important for the conversion of high energy pulses. The presented optical scheme also allows one to synthesize femtosecond vortices with opposite topological charges by simply reversing the handedness of the input polarization with the first quarter-wave plate.

In our experiments we used the output beam of a Coherent Mira 900 laser oscillator with the central wavelength at 800 nm. The pulse duration before the converter (based on measurements with a Femtochrome FR-103 autocorrelator) was 120 fs full width at half maximum (FWHM). After the converter we observed a moderate temporal pulse broadening to ~150 fs FWHM, which can be corrected, if necessary, by pre-chirping the input pulses.

Figure 2 shows the experimental setup used to determine the topological charge of the generated femtosecond vortices. This simple non-interferometric technique was earlier used for cw beams [13]. Here we show that it can also be applied for analysis of femtosecond pulses. It is based on the astigmatic transformation of the incident vortex beam with a cylindrical lens. The resulting intensity distribution of a vortex beam in the focal plane of a cylindrical lens is featured by the presence of tilted dark bands. Their number determines the absolute value of the topological charge, whereas the tilt $\alpha$ (see Fig. 2 and 4) gives its sign.

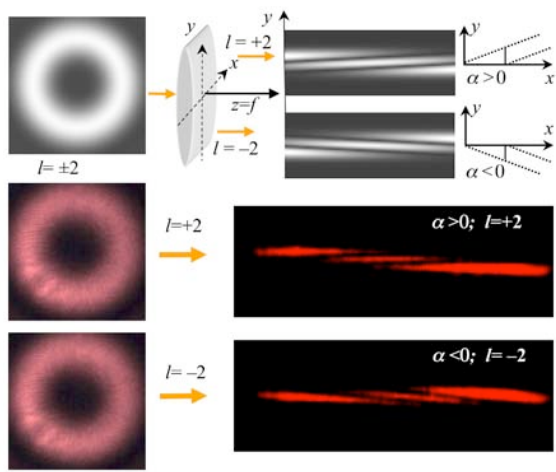

**Fig. 2.** Diagnostics of femtosecond vortex beam after the converter in Fig. 1: evidence of its singular nature and determination of the topological charge. Top: simulated behavior of a double-charge vortex at the focus of a positive cylindrical lens. Bottom: experimentally obtained images of femtosecond double-charge vortices 0.5 m after the converter (left) and their focal caustics (right). Horizontal size of all experimental images is 4.2 mm.

Similar approach to that used to form double-charge vortices has been applied to generate single-charge femtosecond vortex pulses, as shown in Fig. 3. If the input Gaussian beam is linearly x-polarized, then after the crystal we observe a superposition of two vector states of the field: i) a non-vortex y-polarized beam and ii) an optical multiple vortex state, whose polarization is parallel to that of the incident light (i.e., x-polarized). An expression for the vortex state is given by [12]:

$$E_x = \frac{1}{2}\left(G_o + G_e - \left(\frac{w_0^2}{uv}(\xi_o G_o - \xi_e G_e) + G_o - G_e\right)\cos 2\varphi\right) \quad (2)$$

According to Eq. (2), the lines of the field's zeros (i.e., vortices) lie in the four half-planes intersecting at the crystal's axis. These half-planes are oriented at an angle $\varphi = \pi(2m+1)/4$, $m = 0, 1, 2, 3$; to the beam polarization plane. The odd and even values of the integer $m$ correspond to vortices with the opposite topological charges. The positions of the vortices in the half-planes are given by the solutions of the equation $G_o + G_e = 0$. A single vortex closest to the axis could be selected by tilting the crystal by an angle $\theta \approx \sqrt{\lambda/(2z(n_o - n_e))}$ in one of these half-planes. Based on the analysis presented in [12], the maximum achievable ratio of the optical powers in the single-charge vortex beam and the non-vortex beam is 3:1 giving a maximum of 75% conversion efficiency (Fig. 3). To characterize single charge vortices we again employed the astigmatic transformation with a cylindrical lens (Fig.4).

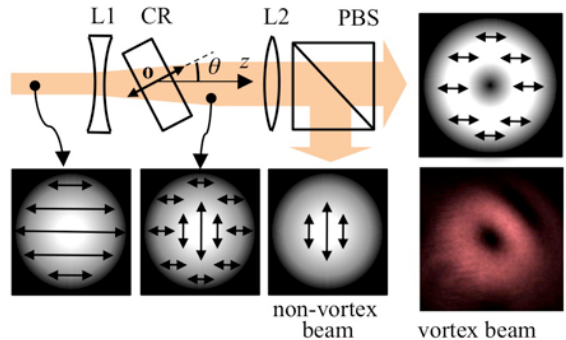

**Fig. 3.** Experimental set-up for the generation of double-charge femtosecond vortex pulses (the notations are as in Fig.1). **o** denotes the optical axis of the crystal. Bottom right: CCD image (4.2 x 4.2 mm$^2$) of a single-charge vortex recorded ~5 m after the converter.

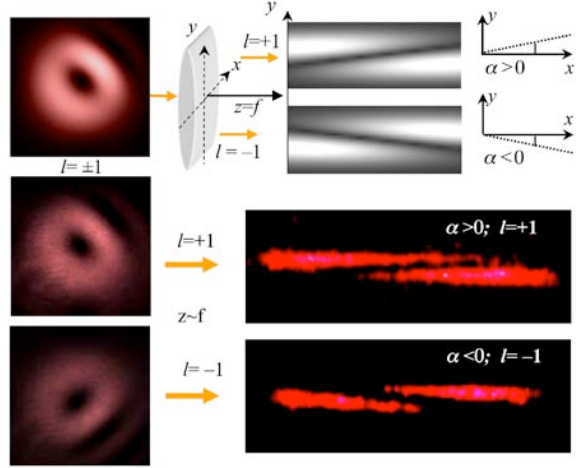

**Fig. 4.** Diagnostics of femtosecond vortex beam after the converter in Fig. 3: evidence of its singular nature and determination of the topological charge. Top: simulated behavior of a single-charge vortex at the focus of a positive cylindrical lens. Bottom: experimental images of femtosecond single-charge vortices 0.5m after the converter (left) and their focal caustics for the opposite topological charges (right). Horizontal size of all experimental images is 4.2 mm.

The high 75% and 50% conversion rates for single- and double-charge vortices, respectively, have been confirmed experimentally using the same converter based on the Galilean telescope comprised of a -50 mm and a +100 mm lens (i.e., L1 and L2 in Fig. 1 and Fig. 2). The input beam diameters giving the maximum conversion efficiency for single- and double-charge vortex beams were ~1.5 mm and ~3.0 mm, respectively.

In summary, we have presented a simple and cost effective optical design for the generation of femtosecond optical vortices based on light propagation inside uniaxial birefringent crystal. The simplicity and adaptability, high conversion efficiency, inherent achromaticity, and suitability for high power applications make this converter useful for a wide range of ultrafast intense-field experiments and laser micromachining applications.

We acknowledge financial support from the National Health and Medical Research Council of Australia and the Australian Research Council.